\documentclass[11pt]{article}
\usepackage{epsfig}

\newcommand{\comment}[1]{}

\def \lket {|}
\def \rket {\rangle}
\def \lbra {\langle}
\def \rbra {|}
\def \zv{\overrightarrow{0}}
\newcommand{\ket}[1]{\lket #1\rket}
\newcommand{\bra}[1]{\lbra #1\rbra}
\def\H {{\cal H}}

\newtheorem{Theorem}{Theorem}
\newtheorem{Lemma}{Lemma}
\newtheorem{Corollary}{Corollary}
\newtheorem{Claim}{Claim}
\newcommand{\proof}{\noindent {\bf Proof: }}
\newcommand{\qed}{\nobreak \ifvmode \relax \else
      \ifdim\lastskip<1.5em \hskip-\lastskip
      \hskip1.5em plus0em minus0.5em \fi \nobreak
      \vrule height0.75em width0.5em depth0.25em\fi}

\begin{document}
\title{A nearly optimal discrete query quantum algorithm 
for evaluating NAND formulas}
\date{}

\author{Andris Ambainis\thanks{Department of Combinatorics and Optimization
and IQC, University of Waterloo, Canada, {\tt ambainis@math.uwaterloo.ca}.
Supported by NSERC, CIAR, ARO/DTO, MITACS and IQC.} }
\maketitle

\abstract{We present an $O(\sqrt{N})$ discrete query quantum algorithm
for evaluating balanced binary NAND formulas and an 
$O(N^{\frac{1}{2}+O(\frac{1}{\sqrt{\log N}})})$
discrete query quantum algorithm for evaluating arbitrary binary NAND 
formulas.}

\section{Introduction}

One of two most famous quantum algorithms is Grover's search \cite{Grover}
which solves a generic problem of exhaustive search among $N$ possibilities
in $O(\sqrt{N})$ steps. This provides a quadratic speedup over the naive
classical algorithm for a variety of search problems \cite{Ambainis-survey}.

Grover's algorithm can be re-cast as computing OR of $N$ bits $x_1, \ldots, x_N$,
with $O(\sqrt{N})$ queries to a black box storing the values $x_1, \ldots, x_N$.
Then, a natural generalization of this problem is computing the value 
of an AND-OR formula of $x_1, \ldots, x_N$. 

This problem can be viewed as a black-box model for determining 
the winner in a 2-player game (such as chess) if both
players play their optimal strategies. 
In this case, the game can be represented by a game tree consisting
of possible positions. The leaves of a tree correspond to the possible end
positions of the game. Each of them contains a variable $x_i$, with $x_i=1$
if the first player wins and $x_i=0$ otherwise. Internal nodes corresponding
to positions which the first player makes the next move contain a value
that is OR of the values of their children. (The first player wins if he has a move 
that leads to a position from which he can win.) Internal nodes for 
which the second player makes the next move contain a value
that is AND of the values of their children. (The first player wins if he wins
for any possible move of the second player.)

The question is: assuming we have no further information about the game 
beyond the position tree, how many of the variables $x_i$ do we have
to examine to determine whether the first player has a winning strategy?
This problem has been studied in both classical \cite{Santha,SW,Snir} and
quantum \cite{Ambainis-lb,Ambainis-var,BCW,HMW} context.
 
Throughout this paper, we will assume that the formula is read-once 
(every leaf contains a different variable). There are two main cases that have been
studied in the quantum case.

The first case is when the formula is of a constant depth $d$. 
If the formula is balanced (which is the most commonly studied case),
then even levels contain ORs of $N^{1/d}$ variables and odd levels
contain ANDs of $N^{1/d}$ variables. In this case, $\Theta(\sqrt{N})$
quantum queries are both sufficient \cite{BCW,HMW} and necessary \cite{Ambainis-lb}.
Since any randomized algorithm requires $\Omega(N)$ queries in this case, 
we still achieve a quadratic speedup over the best classical algorithm.
For arbitrary formulas of depth $d$, $O(\sqrt{N}\log^{d-1} N)$ queries
suffice \cite{Ambainis-var}.
This is almost tight, as Barnum and Saks \cite{BS} have shown that
$\Omega(\sqrt{N})$ queries are necessary to evaluate any AND-OR formula of any depth.

The second case is when, instead of a constant depth, we have a constant fan-out.
This case has been much harder and, until a few months ago, there has been no 
progress on it at all. If we restrict to binary AND-OR trees, the 
classical complexity of computing the value of a balanced binary AND-OR tree
is $\Theta(N^{.754...})$ \cite{Santha,SW,Snir} and there was no better quantum
algorithm known.

In a breakthrough result, Farhi et al. \cite{FGG} showed that  
the value of a balanced binary NAND tree can be computed in $O(\sqrt{N})$ quantum time
in an unconventional continuous-time Hamiltonian query model of \cite{FG,Mochon}.
(Because of De Morgan's laws, computing the value of an AND-OR tree is equivalent
to computing the value of a NAND tree.) 
Using a standard reduction between continuous time and discrete time 
quantum computation \cite{Cleve}, this yields an $O(N^{1/2+\epsilon})$ query
quantum algorithm in the standard discrete time quantum query model, for any
$\epsilon>0$. (The big-O constant deteriorates, as the $\epsilon$ decreases.)

Soon after, Childs et al. \cite{Childs} extended the result of \cite{FGG} to
computing the value of an arbitrary binary NAND tree of depth in time $O(\sqrt{Nd})$ in
the continuous-time Hamiltonian query model and with $O(N^{1/2+\epsilon})$ queries
in the discrete-time query model.

In this paper, we improve over \cite{FGG} and \cite{Childs} by giving a better
discrete time quantum query algorithms for both balanced and general NAND trees.
Namely, we give
\begin{enumerate}
\item
An $O(\sqrt{N})$ query quantum algorithm for evaluating balanced binary NAND formulas,
which is optimal up to a constant factor.
\item
An $O(\sqrt{Nd})$ query quantum algorithm for evaluating arbitrary binary NAND formulas of 
depth $d$.
\item
An $O(N^{\frac{1}{2}+O(\frac{1}{\sqrt{\log N}})})$ 
query quantum algorithm for evaluating arbitrary binary NAND formulas of 
any depth.
\end{enumerate}
All of our algorithms are designed directly in the discrete quantum query model
and do not incur the overhead from converting from continuous to discrete time.

Besides better running time, our algorithms provide a new perspective for
understanding the quantum algorithms for this problem.
When the breakthrough algorithm of \cite{FGG} appeared, its ideas seemed to
be very different from anything known before. 
Our new algorithm and its analysis show intricate connections 
to the previous work on quantum search. 

Although its technical details are complex, the main intuition is
the same as in Grover's search \cite{Grover} and its "two reflections"
analysis \cite{Aha99} which views the Grover's algorithm as a sequence
of reflections in two-dimensional space against two different axes. 
The idea of "two reflections" has come up in quantum algorithms over 
and over. For example, the element distinctness algorithm 
of \cite{Ambainis-distinct}, designed by different methods, was re-cast
in the form of two reflections by Szegedy \cite{Szegedy}.
In this paper, we show that the NAND-tree algorithms can be viewed as
another instance of "two-reflections", with the reflections designed,
using the structure of the NAND tree. 

\section{Preliminaries}

\subsection{Quantum query model}

We work in the standard discrete time quantum 
query model \cite{Ambainis-survey,BWSurvey}.
In this model, the input bits can be accessed by queries to an oracle $X$
and the complexity of $f$ is the number of queries needed to compute $f$.
A quantum computation with $T$ queries
is just a sequence of unitary transformations
\[ V_0\rightarrow O\rightarrow V_1\rightarrow O\rightarrow\ldots
\rightarrow V_{T-1}\rightarrow O\rightarrow V_T.\]

The $V_j$'s can be arbitrary unitary transformations that do not depend
on the input bits $x_1, \ldots, x_N$. 
The $O$'s are query (oracle) transformations
which depend on $x_1, \ldots, x_N$.
To define $O$, we represent basis states as $|i, z\rangle$ where
$i\in\{0, 1, \ldots, N\}$. The query transformation $O_x$ 
(where $x=(x_1, \ldots, x_N)$) maps $\ket{0, z}$ to $\ket{0, z}$ and 
$\ket{i, z}$ to $(-1)^{x_i}\ket{i, z}$ for $i\in\{1, ..., N\}$
(i.e., we change phase depending on $x_i$, unless $i=0$ in which case we do
nothing).

The computation starts with a state $|0\rangle$. 
Then, we apply $V_0$, $O_x$, $\ldots$, $O_x$, $V_T$ 
and measure the final state.
The result of the computation is the rightmost bit of the result of
the measurement.
A quantum algorithm computes a function $f(x_1, \ldots, x_N)$ if,
for any $x_1, \ldots, x_N\in\{0, 1\}$, the probability that the result
of the measurement is equal to $f(x_1, \ldots, x_N)$ is at least 2/3. 

We will describe our algorithm in a high level language but it can be
translated into a sequence of transformations of this form.

\subsection{Phase estimation}

In our algorithm, we use phase estimation \cite{CEMM}.
Assume that we are given a black box performing a unitary transformation
$U$ and a state $\ket{\psi}$ which is an eigenstate of $U$: 
$U\ket{\psi}=e^{i\theta} \ket{\psi}$.
Our goal is to obtain an estimate $\tilde{\theta}$ such that
$|\tilde{\theta}-\theta|<\delta$ with probability at least $1-\epsilon$.
The algorithm for phase estimation by \cite{CEMM} solves this problem 
by invoking $U$ $O(\frac{1}{\delta\epsilon})$ times.

If the input to this algorithm is a state $\ket{\psi}$ that is a linear 
combination of different eigenstates: $\ket{\psi}=\sum_j \alpha_j \ket{\psi_j}$
with $U_i\ket{\psi_j}=e^{i \theta_j} \ket{\psi_j}$, then the algorithm
works as if the input was a probabilistic combination of $\ket{\psi_j}$
with probabilities $|\alpha_j|^2$.

\section{Summary of results and methods}

\subsection{Results}

Let $T$ be a read-once binary NAND formula involving 
variables $x_1, x_2$, $\ldots$, $x_{N}$. 
We can represent $T$ by a tree that have variables $x_1, \ldots, x_N$
at the leaves and NAND gates at the internal nodes.
Let $d$ be the depth of $T$.
We have

\begin{Theorem}
\label{thm:alg}
\begin{enumerate}
\item
If $T$ is the complete binary tree, then $T(x_1, \ldots, x_N)$ can
be computed with $O(\sqrt{N})$ quantum queries.
\item
For any binary tree $T$, $T(x_1, \ldots, x_N)$ can
be computed with $O(\sqrt{dN})$ quantum queries.
\end{enumerate}
\end{Theorem}

We refer to the first part of the theorem as the {\em balanced case}
and to the second part as the {\em general case}.

Bshouty et al. \cite{BCE} have shown

\begin{Theorem}
\label{thm:bce}
\cite{BCE}
For any NAND formula $T$ of size $S$, 
there exists a NAND formula $T'$ of size $S'=O(S^{1+O(\frac{1}{\sqrt{\log S}})})$
and depth $d=O(S^{O(\frac{1}{\sqrt{\log S}})})$ such that $T'=T$.
\end{Theorem}

This theorem follows by substituting $k=2^{\frac{1}{\sqrt{\log S}}}$ into
Theorem 6 of \cite{BCE}.
By combining Theorems \ref{thm:alg} and \ref{thm:bce}, we have

\begin{Corollary}
 For any $T$, $T(x_1, \ldots, x_N)$ can
be computed with $O(N^{\frac{1}{2}+O(\frac{1}{\sqrt{\log N}})})$ quantum queries.
\end{Corollary}

If the formula $T$ is not read once, the number of variables $N$ is replaced by the
size of the formula $S$. This gives us

\begin{Corollary}
\label{cor:laplante}
If $T(x_1, \ldots, x_N)$ is computable by a NAND formula of size $S$, $T$ is 
computable by a quantum query algorithm with
$O(S^{\frac{1}{2}+O(\frac{1}{\sqrt{\log S}})})$ queries.
\end{Corollary}

The link between quantum query complexity and formula size was first noticed by
Laplante et al. \cite{LLS} who observed that, whenever quantum adversary lower
bound method of \cite{Ambainis-lb} 
gives a lower bound of $\Omega(M)$ for quantum query algorithms,
it also gives a lower bound of $\Omega(M^2)$ for formula size.
Based on that, they conjectured that any Boolean function with formula size $M^2$
has a quantum query algorithm with $O(\sqrt{M})$ queries. 
The results in \cite{Childs} and this paper show that it is indeed possible
to transform an arbitrary NAND formula into a quantum query algorithm,
with almost a quadratic relation between formula size and the number of queries.

\subsection{The algorithm}
\label{sec:alg}

Our algorithm is the same for both parts of Theorem \ref{thm:main}.
Without the loss of generality, assume that all leafs are at an even 
distance from the root. (If there is a leaf $l$ at an odd depth, create
two new vertices $v_1, v_2$ and connect them to 
$l$, making $l$ an internal node.
$v_1, v_2$ are now leaves at an even depth.
If $x_i$ is the variable that used to be at the leaf $l$,
replace it by two new variables at leaves $v_1, v_2$ and make both
of those equal to $NOT x_i$. Then, the NAND of those two
variables at the vertex $l$ will evaluate to $x_i$.)

\begin{figure}[ht]
\center{\epsfxsize=4in\epsfbox{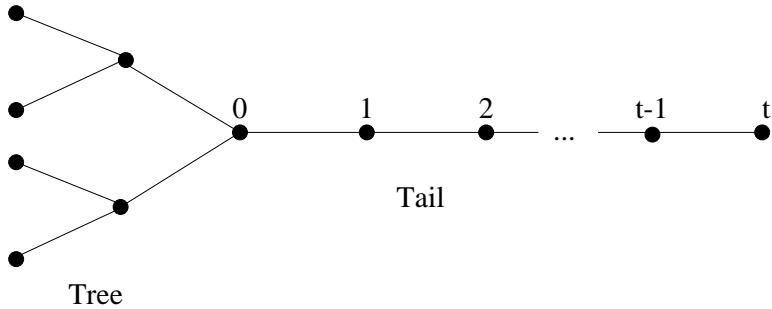}}
\caption{A tree $T$, augmented by a tail.}
\label{fig:tail}
\end{figure}

We augment the tree $T$ by a ``tail" of an even length
$t$ where $t=2 \lceil \sqrt{N} \rceil$ in the balanced case and
$t=2\lceil \sqrt{Nd}\rceil$ in the general case.
The tail is a path that starts at the root of $T$ and then goes through
$t$ newly created vertices.
$T'$ denotes the tree $T$, augmented by the tail (see 
Figure \ref{fig:tail}).
 
Our state space $\H$ will be spanned by basis
states $\ket{v}$ corresponding to vertices of $T'$.
We use $\ket{i}$ (for $i=0, \ldots, t$) to denote
the basis state corresponding to
the $i^{\rm th}$ vertex in the tail of $T'$.
$\ket{0}$ corresponds to the root of $T$.

We define a Hermitian matrix $H$ as follows:
\begin{enumerate}
\item
If $pc$ is an edge in $T$ from a parent $p$ to a child $c$, then
$H_{pc}=H_{cp}=\sqrt[4]{\frac{m_p}{2m_c}}$ if $p$ is at an odd
level and $H_{pc}=H_{cp}=\sqrt[4]{\frac{2m_c}{m_p}}$ if $p$ is at an even level.
(When $T$ is the complete balanced tree, this becomes $H_{cp}=H_{pc}=1$.)
\item
If $uv$ is an edge in the tail, then $H_{uv}=H_{vu}=1$.
\item
If $uv$ is not an edge, then $H_{uv}=H_{vu}=0$.
\end{enumerate}

Then, $H$ is a Hermitian operator acting on $\H$.
Let $S_{H, 0}$ be the 0-eigenspace of $H$, $S_{H, 1}=(S_{H, 0})^{\perp}$ 
and let $U_1$ be defined by
$U_1\ket{\psi}=\ket{\psi}$ for $\ket{\psi}\in S_{H, 0}$ and 
$U_1\ket{\psi}=-\ket{\psi}$ for $\ket{\psi}\in S_{H, 1}$.

Let $U_2$ be defined by $U_2\ket{\psi}=-\ket{\psi}$ if $\ket{\psi}$ belongs
to the subspace $S_{x, 1}$ spanned by basis states $\ket{v}$ that correspond 
to leaves containing variables $x_i=1$ and $U_2\ket{\psi}=\ket{\psi}$ if $\ket{\psi}$ belongs
to the subspace $S_{x, 0}$ spanned by all other basis states $\ket{v}$.

$U_2$ can be implemented with one query $O$. (It is essentially the query
transformation $O_x$, with basis states labelled in a different way.)
$U_1$ is independent of $x_1, \ldots, x_N$ and can be implemented without
using the query transformation $O$.

Let $\ket{\psi_{start}}=\sum_{i=0}^{t/2} \ket{2i}$. 
(This is the starting state for the continuous time algorithm of Childs et al. \cite{Childs}.)
Let $\ket{\psi'_{start}}=P_{S_{H, 0}} \ket{\psi_{start}}$ 
and let $\ket{\psi''_{start}}=\frac{\ket{\psi'_{start}}}{\|\psi'_{start}\|}$.

\begin{Theorem}
\label{thm:main}
\begin{enumerate}
\item
If $T$ evaluates to 0, there is a state $\ket{\psi_0}$ such that
$U_2 U_1 \ket{\psi_0}=\ket{\psi_0}$ and 
$|\lbra \psi_0 | \psi''_{start}\rket|^2\geq c$ for some constant $c>0$.
\item
If $T$ evaluates to 1, then, for any eigenstate $\ket{\psi_0}$ of $U_2U_1$ 
which is not orthogonal to $\ket{\psi''_{start}}$, the corresponding
eigenvalue of $U_2 U_1$ is $e^{i\theta}$, with
$\theta=\Omega(\frac{1}{\sqrt{Nd}})$ for any $T$ and 
$\theta=\Omega(\frac{1}{\sqrt{N}})$ when $T$ is the complete balanced tree. 
\end{enumerate}
\end{Theorem}

We can distinguish the two cases by running the eigenvalue estimation
for $U_2 U_1$, with $\ket{\psi''_{start}}$ as the starting state, 
precision $\delta=\frac{\theta_{min}}{2}$ where $\theta_{min}$ 
is the lower bound on $\theta$ from the second part of Theorem \ref{thm:main}
($\theta_{min}=\Theta(\frac{1}{\sqrt{Nd}})$ or $\theta_{min}=\Theta(\frac{1}{\sqrt{N}})$) 
and error probability $\epsilon\leq \frac{c}{3}$. 
In the first case, with probability 
$|\lbra \psi_0 | \psi''_{start}\rket|^2\geq c$,
we get the same answer as if the input to eigenvalue estimation
was $\ket{\psi_0}$. Since the correct eigenvalue is 0, 
this means that we get an answer $\tilde{\theta}<\frac{\theta_{min}}{2}$ 
with probability at least $(1-\epsilon)c$.

In the second case, if we write out $\ket{\psi''_{start}}$ as 
a linear combination of eigenvectors of $U_2 U_1$, 
all of those eigenvectors have eigenvalues that are $e^{i\theta}$, 
$\theta>\theta_{min}$. Therefore, the probability
of the eigenvalue estimation outputting an estimate 
$\tilde{\theta}<\theta_{min}-\delta=\frac{\theta_{min}}{2}$ is
at most $\epsilon$. 

By our choice of $\epsilon$, we have $(1-\epsilon)c>\epsilon$. 
We can distinguish the two cases with arbitrarily high probability,
by repeating the eigenvalue estimation $C$ times, for a sufficiently
large constant $C$.

\subsection{Proof overview}

The first part of Theorem \ref{thm:main} is proven by constructing 
the state $\ket{\psi_0}$. For the second part, we show that the entire
state-space $\H$ can be expressed as a direct sum of one-dimensional
and two-dimensional subspaces, with each subspace being mapped to itself
by $U_1$ and $U_2$. Each one-dimensional subspace consists of all multiples
of some state $\ket{\psi}$, with $U_1\ket{\psi}$ and $U_2\ket{\psi}$ 
being either $\ket{\psi}$ or $-\ket{\psi}$. 
Therefore, we either have $U_2U_1\ket{\psi}=\ket{\psi}$ or
$U_2U_1\ket{\psi}=-\ket{\psi}$.
We show that, if $U_2U_1\ket{\psi}=\ket{\psi}$, then
$\ket{\psi}$ is orthogonal to the starting state
$\ket{\psi''_{start}}$ and, therefore, has no effect on the algorithm.

For two-dimensional subspaces, we show that each of them
has an orthonormal basis $\ket{\psi_{11}}$, $\ket{\psi_{12}}$ such that
$U_1\ket{\psi_{11}}=\ket{\psi_{11}}$ and $U_1\ket{\psi_{12}}=-\ket{\psi_{12}}$
and another orthonormal basis $\ket{\psi_{21}}$, $\ket{\psi_{22}}$ such that
$U_2\ket{\psi_{21}}=\ket{\psi_{21}}$ and $U_2\ket{\psi_{22}}=-\ket{\psi_{22}}$.
Then, on this two-dimensional subspace,
$U_2U_1$ is a product of two reflections, one w.r.t. $\ket{\psi_{11}}$
and one w.r.t. $\ket{\psi_{21}}$. As in "two reflections" analysis \cite{Aha99}
of Grover's search, a product of two reflections in a two-dimensional
plane is a rotation of plane by $2\beta$, where $\beta$ is the angle between
$\ket{\psi_{11}}$ and $\ket{\psi_{21}}$. A rotation of the plane by $2\beta$
has eigenvalues $e^{\pm i\beta}$. Therefore, we need to lower-bound $\beta$.

Since $\ket{\psi_{21}}$ and $\ket{\psi_{22}}$ are orthogonal, 
the angle between $\ket{\psi_{11}}$ and $\ket{\psi_{22}}$ is
$\frac{\pi}{2}-\beta$. Therefore, $|\lbra \psi_{22} \ket{\psi_{11}}|=\sin \beta$. 
Since $\ket{\psi_{22}}$ belongs to $S_{x, 1}$ and $\ket{\psi_{11}}$ belongs
to $S_{H, 0}$, we have 
\[ |\lbra \psi_{22} \ket{\psi_{11}}|=\| P_{S_{x, 1}} \ket{\psi_{11}} \| \geq
\min_{\ket{\psi}\in S_{H, 0}} \| P_{S_{x, 1}} \ket{\psi} \| .\]
Therefore, to lower-bound $\sin \beta$ and $\beta$, it suffices to lower-bound
the minimum of $\| P_{S_{x, 1}} \ket{\psi} \|$ for $\ket{\psi}\in S_{H, 0}$.
We do that by an induction over the depth of the tree. 

\section{Notation}

In this section, we summarize the main notation used in this paper:

{\bf Trees.} 
$T$ is the tree which we are evaluating. $T'$ is the tree $T$ with
the tail attached to it. $T_v$ is the subtree of $T$ rooted at $v$.
We also use $T$ (or $T_v$) to denote the Boolean function defined by evaluating
the NAND tree $T$ (or $T_v$).

$m_v$ and $d_v$ denote the number of leaves and the depth of $T_v$.
$r$ denotes the root of $T$. Thus, $T_r=T$.

{\bf Matrices.}
$H$ is the weighted version of the adjacency matrix of $T'$, defined
in section \ref{sec:alg}.
$H_v$ is the restriction of $H$ to rows and columns in $T_v$.

{\bf Subspaces.}
$S_{H, 0}$ is the eigenspace of $H$ with the eigenvalue 0. 
$S_{H, 1}$ is the orthogonal complement of $S_{H, 0}$:
$S_{H, 1}=(S_{H, 0})^{\perp}$.
$S_{v, 0}$ denotes the 0-eigenspace of $H_v$.
$S'_{H, 0}$ and $S'_{v, 0}$ are subspaces of $S_{H, 0}$ and $S_{v, 0}$,
defined in section \ref{sec:t1}.

$S_{x, 1}$ is the subspace 
spanned by $\ket{v}$, for all leaves $v$ 
that correspond to a variable $x_i=1$. $S_{x, 0}$ is the subspace spanned
by all other $\ket{v}$ (for $v$ that are either 
leaves corresponding to $x_i=0$
or non-leaves). 

{\bf Unitary transformations}
$U_1$ is defined by $U_1\ket{\psi}=\ket{\psi}$ for 
$\ket{\psi}\in S_{H, 0}$ and $U_1\ket{\psi}=-\ket{\psi}$ for  
$\ket{\psi}\in S_{H, 1}$.
$U_2$ is the query transformation. It can be equivalently
described by defining $U_2\ket{\psi}=\ket{\psi}$ for $\ket{\psi}\in S_{x, 0}$
and $U_2\ket{\psi}=-\ket{\psi}$ for $\ket{\psi}\in S_{x, 1}$.

\section{Structure of minimal certificates of $T_v$}
\label{sec:structure}

Let $C$ be a minimal certificate of $T_v=0$. We would like to determine 
the structure of $C$. Let $z_1$ and $z_2$ be
the two children of $v$ and $y_1$, $y_2$ ($y_3$, $y_4$) be the 
children of $z_1$ ($z_2$, respectively).
For $T_v=0$, we need to have $T_{z_1}=T_{z_2}=1$ which is equivalent
to at least one of $T_{y_1}$ and $T_{y_2}$ and at least one
of $T_{y_3}$ and $T_{y_4}$ evaluates to 0.
Thus, a minimal 0-certificate for $T_v$ consists of a minimal 
0-certificate for one of $T_{y_1}=0$ and $T_{y_2}=0$ and a minimal 
0-certificate for one of $T_{y_3}=0$ and $T_{y_4}=0$.
Each of those 0-certificates can be decomposed in a similar way.

\begin{figure}[h]
\center{\epsfxsize=4in \epsfbox{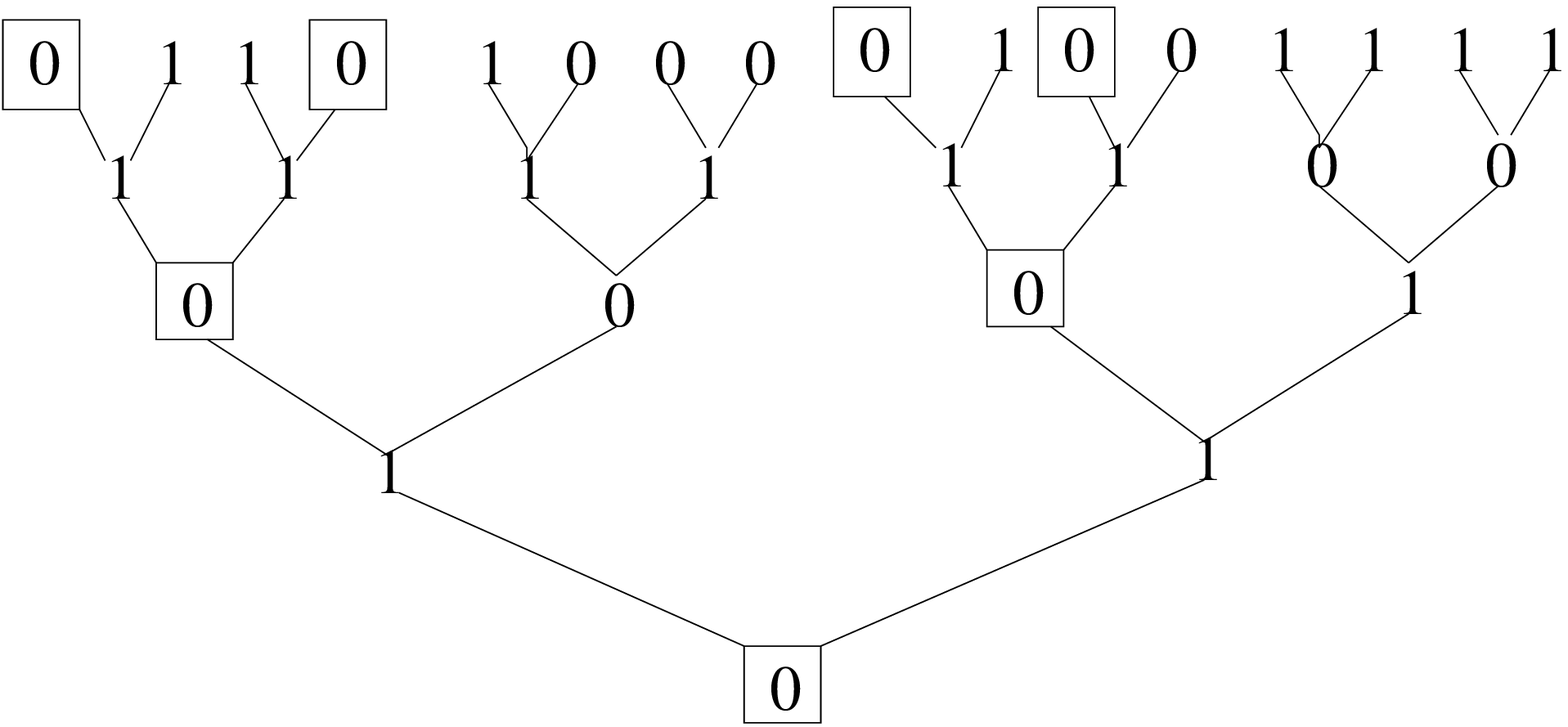}}
\caption{An extended certificate for $T_v=0$.}
\label{fig:cert}
\end{figure}

We now define an extended minimal certificate for $T_v=0$ to consist
of $v$, an extended minimal certificate for one of $T_{y_1}=0$ 
and $T_{y_2}=0$ and an extended minimal 
0-certificate for one of $T_{y_3}=0$ and $T_{y_4}=0$.
Intuitively, an extended minimal certificate is a minimal certificate,
augmented by non-leaf vertices that must evaluate to 0 on this certificate.
We can show

\begin{Lemma}
\label{lem:structure}
Let $C$ be an extended minimal certificate. 
If a non-leaf vertex $w$ belongs to $C$, $z_1$ and $z_2$ are
the two children of $w$ and $y_1$, $y_2$ ($y_3$, $y_4$) are the 
children of $z_1$ ($z_2$, respectively), then exactly one of $y_1, y_2$
and exactly one of $y_3, y_4$ belongs to $C$.
\end{Lemma}

\proof
In appendix \ref{app:structure}.
\qed

We show an example of an extended certificate for $T_v=0$
in figure \ref{fig:cert}. The vertices that
belong to the extended certificate are shown by squares. 

For each extended minimal certificate $C$ of $T_v=0$, we can define a state
$\ket{\psi_C}$ that has non-zero amplitudes only in the vertices of $C$,
in a following way:
\begin{enumerate}
\item
Decompose $C$ as $C=C_{y_i}\cup C_{y_j}\cup\{v\}$,
where $C_{y_i}$ is an extended certificate for $T_{y_i}=0$, $i\in\{1, 2\}$ and
$C_{y_j}$ is an extended certificate for $T_{y_j}=0$, $j\in\{3, 4\}$.
\item
Construct $\ket{\psi_{C_{y_i}}}$ and $\ket{\psi_{C_{y_j}}}$ inductively
and define
\begin{equation}
\label{eq:pc} 
\ket{\psi_C}=\ket{v}-\frac{\sqrt[4]{m_v}}{\sqrt[4]{4m_{y_i}}}
\ket{\psi_{C_{y_i}}}
-\frac{\sqrt[4]{m_v}}{\sqrt[4]{4m_{y_j}}}\ket{\psi_{C_{y_j}}} .
\end{equation}
\end{enumerate}

\begin{Lemma}
\label{lem:0eigen}
Let $C$ be an extended minimal certificate for $T_v=0$. Then,
\[ H_v\ket{\psi_C}=0.\]
\end{Lemma}

\proof
In appendix \ref{app:structure}.
\qed

\begin{Lemma}
\label{lem:norm}
\begin{enumerate}
\item[(a)]
If $T$ is balanced, 
$\|\psi_{C_v}\|^2 \leq 2 \sqrt{m_v}-1$.
\item[(b)]
For any $T$, $\|\psi_{C_v}\|^2 \leq 2\sqrt{m_v d_v}$.
\end{enumerate}
\end{Lemma}

\proof
In appendix \ref{app:structure}.
\qed

\section{Proof of Theorem 1: $T=0$ case}

Let $C$ be an extended minimal 0-certificate
of $T_r$ where $r$ is the root of the tree and let
$\ket{\psi_C}$ be the corresponding state (defined so that the
amplitude $\alpha_r$ of the root is 1).  
We define  
$\ket{\psi_0}=\ket{\psi_C}+\sum_{i=1}^{t/2} (-1)^{i} \ket{2i}$.
Let $\ket{\psi'_0}$ be the corresponding normalized
state: $\ket{\psi'_0}=\frac{\ket{\psi_0}}{\|\psi_0\|}$.

We claim that $U_2 U_1\ket{\psi_0}=\ket{\psi_0}$.
This follows from $U_2\ket{\psi}=\ket{\psi}$ (which is true,
because $\ket{\psi_C}$ and $\ket{\psi_0}$ are only non-zero on the vertices that
belong to the extended certificate $C$ and $x_i=0$ for all variables
$x_i$ at the leaves that belong to a certificate $C$)
and $U_1\ket{\psi_0}=\ket{\psi_0}$ (which follows from the next lemma).

\begin{Lemma}
\label{lem:0eigen1}
\[ H\ket{\psi_0}=0.\]
\end{Lemma}

\proof
It suffices to show that, for every $u$, the amplitude of $u$
in $H\ket{\psi_0}$ is 0. For vertices in the tree $T$, their amplitudes in
$H\ket{\psi_0}$ are the same as their amplitudes in $H_r\ket{\psi_C}$ and,
by Lemma \ref{lem:0eigen}, $H_r\ket{\psi_C}=0$.

For vertices $j$ in the tail, the amplitude of $j$ in $H\ket{\psi_0}$
is the sum of the amplitudes of its two neighbors of $j-1$ and $j+1$
in $\ket{\psi_0}$. If $j=2i$ is even, then both $2i-1$ and $2i+1$ have amplitudes 0
and their sum is 0. If $j=2i+1$ is odd, then one of $2i$ and $2i+2$ has amplitude 1
and the other has amplitude -1, resulting in the sum of amplitudes being 0.
\qed

To complete the proof of the first part of Theorem \ref{thm:main}, we show

\begin{Lemma}
If $t>\sqrt{N}$ (for the balanced case) or $t>\sqrt{Nd}$ 
(for the unbalanced case), then
\[ \bra{\psi'_0} \psi''_{start}\rket \geq \frac{1}{\sqrt{5}} .\]
\end{Lemma}

\proof
We show the proof for the balanced case. (For the unbalanced case,
just replace $N$ by $Nd$ everywhere.)

Since $\ket{\psi'_0}\in S_{H, 0}$ (by Lemma \ref{lem:0eigen1}),
we have
\[ \bra{\psi'_0} \psi_{start}\rket 
= \bra{\psi'_0} P_{S_{H, 0}} \ket{\psi_{start}} =
\|\psi'_{start}\| \bra{\psi'_0}  \psi''_{start} \rket  \leq
\bra{\psi'_0}  \psi''_{start} \rket .\]
Therefore, it suffices to prove 
$\bra{\psi'_0} \psi_{start}\rket \geq \frac{1}{\sqrt{5}}$.
We have
\[ \bra{\psi'_0} \psi_{start}\rket = \frac{\bra{\psi_0}\psi_{start}\rket
}{\|\psi_0\|} .\]
Since each of the basis states $\ket{2j}$ has amplitude 1 in $\ket{\psi_0}$
and amplitude $\frac{1}{\sqrt{\frac{t}{2}+1}}$ in $\ket{\psi_{start}}$,
we have $\bra{\psi_0}\psi_{start}\rket =\sqrt{\frac{t}{2}+1}$.
We also have
\[ \|\psi_0\|^2 = \|\psi_C\|^2 + \frac{t}{2} \leq 
2 \sqrt{N} + \frac{t}{2} \leq 2.5 t .\]
Therefore,
\[ \frac{\bra{\psi_0}\psi_{start}\rket}{\|\psi_0\|} \geq 
\frac{\sqrt{\frac{t}{2}+1}}{\sqrt{2.5 t}}=\frac{1}{\sqrt{5}} .\]
\qed

$\lbra{\psi'_0} | \psi''_{start}\rket$ can be increased to $1-\epsilon$
by taking $t\geq C\sqrt{N}$ for sufficiently large constant $C$.

\section{Proof of Theorem 1: $T=1$ case}
\label{sec:t1}

\subsection{Overview}
\label{sec:overview1}

We first describe a subset of the 1-eigenstates $\ket{\psi}$ of $U_2 U_1$.
Let $v$ be a vertex of an odd depth $2j+1$ and let
$v_1$ and $v_2$ be the children of $v$.
Assume that $T_{v_1}=T_{v_2}=0$ and let $C_1, C_2$ be
the extended minimal certificates for $T_{v_1}=0$ and $T_{v_2}=0$.
Define $\ket{\psi_{C_1, C_2}}=\sqrt[4]{m_{v_1}} \ket{\psi_{C_1}}-
\sqrt[4]{m_{v_2}} \ket{\psi_{C_2}}$.

\begin{Lemma}
\label{lem:fixed}
$U_2\ket{\psi_{C_1, C_2}}=U_1\ket{\psi_{C_1, C_2}}=\ket{\psi_{C_1, C_2}}$.
\end{Lemma}

\proof
In appendix \ref{app:structure}.
\qed

We define $S'_{H, 0}$ to be the subspace spanned by all $\ket{\psi_{C_1, C_2}}$, for
all possible choices of $v$, $C_1$ and $C_2$.
Observe that $S'_{H, 0}$ is orthogonal to the starting state $\ket{\psi_{start}}$
(since the state $\ket{\psi_{start}}$ only has non-zero amplitude on the root and
in the tail and any of the states $\ket{\psi_{C_1, C_2}}$ always has zero amplitudes there).
Since $S'_{H, 0}\subseteq S_{H, 0}$, $S'_{H, 0}$ is also orthogonal to
$\ket{\psi'_{start}}=P_{S_{H, 0}} \ket{\psi_{start}}$
and $\ket{\psi''_{start}}=\frac{\ket{\psi'_{start}}}{\|\psi'_{start}\|}$.

Theorem \ref{thm:main} follows from the following two lemmas:

\begin{Lemma}
\label{lem:easy}
Let $S$ be a Hilbert space, let
$S_1, S_2$ be two subspaces of $S$ and let $U_i$ ($i\in\{1, 2\}$) 
be the unitary transformation on $S$ defined by $U_i\ket{\psi}=-\ket{\psi}$
for $\ket{\psi}\in S_i$ and $U_i\ket{\psi}=\ket{\psi}$ for
$\ket{\psi}\in (S_i)^{\perp}$. 
Assume that $S_1\cap S_2=\{\zv\}$ and, for any $\ket{\psi}\in S_1$, we have
\[ \|P_{(S_2)^{\perp}} \ket{\psi}\|^2 \geq \epsilon .\]
Then, all eigenvalues of $U_2 U_1$ are of the form $e^{i\theta}$
with $\theta\in[\sqrt{\epsilon}, 2\pi-\sqrt{\epsilon}]$.
\end{Lemma}

\begin{Lemma}
\label{lem:project1}
For any state $\ket{\psi}\in S_{H, 0} \cap (S'_{H, 0})^{\perp}$, 
$\|P_{S_{x, 1}}\ket{\psi}\|^2 \geq c \|\psi\|^2$
for a constant $c$, where $c=\Omega(\frac{1}{N})$ in the balanced case
and $c=\Omega(\frac{1}{Nd})$ in the general case.
\end{Lemma}

Given the two lemmas, the proof of Theorem is completed as follows.
Define $S=(S'_{H, 0})^{\perp}\cap (S_{H, 1}\cap S_{x, 1})^{\perp}$. 
For both $S'_{H, 0}$ and $S_{H, 1}\cap S_{x, 1}$, $U_1$ and $U_2$ map
those subspaces to themselves. Since $U_1$ and $U_2$ are unitary,
this means that $U_1$ and $U_2$ map $S$ to itself, as well.
Combining Lemma \ref{lem:easy} and \ref{lem:project1} implies
that all eigenvalues of $U_2 U_1$ on $S$ are $e^{i\theta}$
with $\theta\in[\delta, 2\pi-\delta]$, with 
$\delta=\Omega(\frac{1}{\sqrt{N}})$ in the balanced case
and $\delta=\Omega(\frac{1}{\sqrt{Nd}})$ in the general case.

Since $S'_{H, 0}$ and $S_{H, 1}$ are both orthogonal to $\ket{\psi''_{start}}$,
the state $\ket{\psi''_{start}}$ belongs to $S$.
Therefore, all eigenvectors of $U_2 U_1$ which are not orthogonal to 
$\ket{\psi''_{start}}$ must lie in $S$ and, therefore,
have eigenvalues $e^{i\theta}$ of the required form.
This completes the proof of the theorem.

Lemma \ref{lem:easy} is fairly similar to previous work.
We give its proof in Appendix \ref{app:easy}.
In the next section, we give the proof of Lemma \ref{lem:project1},
postponing some of the technical details till the appendices.

\subsection{Proof of Lemma \ref{lem:project1}}

This lemma is equivalent to the next one.

\begin{Lemma}
\label{lem:project2}
For any state $\ket{\psi}\in S_{H, 0}$, there exists a state 
$\ket{\psi'}\in S'_{H, 0}$ such that
\[ \|P_{S_{x, 1}}(\ket{\psi}+\ket{\psi'})\|^2 \geq 
c \|\ket{\psi}+\ket{\psi'}\|^2 \]
for a constant $c$, where $c=\Omega(\frac{1}{N})$ in the balanced case
and $c=\Omega(\frac{1}{Nd})$ in the general case.
\end{Lemma}

\proof [of Lemma \ref{lem:project1}]
By construction of $S'_{H, 0}$, any state in $S'_{H, 0}$ 
is orthogonal to $S_{x, 1}$.
Let $\ket{\psi}\in S_{H_0}\cap (S'_{H, 0})^{\perp}$. We use 
Lemma \ref{lem:project2} to obtain $\ket{\psi'}\in S'_{H, 0}$.
Then, we have 
\[ \|P_{S_{x, 1}} \ket{\psi}\|^2 = \|P_{S_{x, 1}}(\psi+\psi')\|^2  \geq 
c \|\psi+\psi'\|^2 = c \|\psi\|^2 + c\|\psi'\|^2 \geq c \|\psi\|^2 ,\]
with the first equality following from $\ket{\psi'}$ being orthogonal 
to $S_{x, 1}$, the second inequality from lemma \ref{lem:project2} and the
the third equality following from $\ket{\psi}$ being orthogonal to $\ket{\psi'}$
(which is true because $\ket{\psi'}\in S'_{H, 0}$ but $\ket{\psi}\in (S'_{H, 0})^{\perp}$). 
\qed

\proof [of Lemma \ref{lem:project2}]
We recall that $S_{v, 0}$ is the 0-eigenspace of $H_v$. 
We define $S'_{v, 0}$ to be the subspace spanned by all
$\ket{\psi_{C_1, C_2}}$ which only have non-zero amplitudes for $\ket{u}$, $u\in T_v$.
Then, we have $S'_{v, 0}\subseteq S_{v, 0}$.

For a state $\ket{\psi}$, we define
\[ L_v(\psi)=\frac{\|\psi\|^2 - K \| P_{S_{x, 1}} \psi\|^2}{|\alpha_v|^2} \]
where $\alpha_v$ is the amplitude of $v$ of $T_k$ in $\ket{\psi}$
and $K$ will be defined later.
\comment{If $k$ is odd, then the root is at an odd depth and its amplitude is 0.
In this case, we express $\ket{\psi}=\ket{\psi_{left}}+\ket{\psi_{right}}$,
with $\ket{\psi_{left}}$ ($\ket{\psi_{right}}$) being a superposition
over the left (right) subtree of $T_k$.
Let $\alpha_{left}$ ($\alpha_{right}$) be the amplitude of the
root of the left subtree (the right subtree) in $\ket{\psi_{left}}$ 
($\ket{\psi_{right}}$). Then, we define
\[ L(\psi)=\frac{\|\psi\|^2 - K \| P_{S_{x, 1}} \psi\|^2}{
|\alpha_{\left}|^2+|\alpha_{\right}|^2} .\]}

The next lemma is slightly different for balanced NAND trees and general
NAND trees. Below is the variant for balanced trees. The counterpart for
general trees is described in appendix \ref{app:recursion-full}.

\begin{Lemma}
\label{lem:project3-short}
Let $K\geq 20N$ and let $v$ be a vertex at depth $2k$.
For any state $\ket{\psi}\in S_{v, 0}$, there exists a state 
$\ket{\psi'}\in S'_{v, 0}$ such that, 
for $\ket{\psi''}=\ket{\psi}+\ket{\psi'}$, we have 
\begin{enumerate}
\item
If $T_k(x_1, \ldots, x_N)=0$, then 
\[ L_v(\psi)\leq a_v, \mbox{~~} 
a_v=\left( 1+ 2\frac{2^{2k}}{K} \right) (2^{k+1}-1) .\]
\item
If $T_k(x_1, \ldots, x_N)=1$, then 
\[ L_v(\psi)\leq  -b_v, \mbox{~~} 
b_v=\left( 1- 2\frac{2^{2k}}{K} \right) \frac{K}{2^k}.\]
\end{enumerate}
\end{Lemma}

Given Lemma \ref{lem:project3-short}, the proof of Lemma \ref{lem:project2}
can be completed as follows. Let $\ket{\psi}$ be a 0-eigenstate of $H$.
Then, we can decompose $\ket{\psi}=\ket{\psi_{tree}}+\ket{\psi_{tail}}$,
with $\ket{\psi_{tree}}$ being a superposition over vertices in $T$ 
(including the root) and $\ket{\psi_{tail}}$ being a superposition
over vertices in the tail. 

Let $\alpha_r$ be the amplitude of the root
in $\ket{\psi}$. In both balanced and general case, we have

\begin{Lemma}
\label{lem:odd}
Let $\ket{\psi}=\sum_u \alpha_u \ket{u}$ be a 0-eigenstate of $H$
(or $H_v$, for some $v$).
Then, any vertex $u$ of an odd depth (or any $u$ in the tail at an odd 
distance from the root) must have $\alpha_u=0$.
\end{Lemma}

\proof
In appendix \ref{app:structure}.
\qed

By this lemma, $H\ket{\psi}=0$ implies that the amplitudes
of the vertices in the tail at an odd distance from the root are 0.
Also, to achieve $H\ket{\psi}=0$, the amplitudes at an even distance from
the root must be $\pm \alpha_r$.
Therefore, $\|\psi_{tail}\|^2= \frac{t}{2} |\alpha_r|^2$.

We have $H_{r} \ket{\psi_{tree}}=0$. Therefore, we can 
apply Lemma \ref{lem:project3-short} with $v=r$ 
and $\ket{\psi_{tree}}$ instead of $\ket{\psi}$, obtaining
a state $\ket{\psi'}\in S'_{r, 0}\subseteq S'_{H, 0}$. 
We let
$\ket{\psi''_{tree}}=\ket{\psi_{tree}}+\ket{\psi'}$ and 
$\ket{\psi''_{all}}=\ket{\psi}+\ket{\psi'}
=\ket{\psi''_{tree}}+\ket{\psi_{tail}}$.
Then, 
\[ \|\psi''_{all}\|^2=\|\psi''_{tree}\|^2 + \frac{t}{2} |\alpha_r|^2. \]
We have $P_{S_{x, 1}}\ket{\psi''_{all}}=P_{S_{x, 1}} \ket{\psi''_{tree}}$.
From Lemma \ref{lem:project3-short}, 
\[ \|\psi''_{tree}\|^2\leq K \|P_{S_{x,1}}\ket{\psi''_{tree}}\| - b_r 
|\alpha_r|^2 . \]
This means that  
\[ \|\psi''_{all}\|^2 + \left(b_r -\frac{t}{2} \right) |\alpha_r|^2 \leq
K \|P_{S_{x,1}}\ket{\psi''_{all}}\| .\]
If $t=2\lceil \sqrt{N}\rceil$, we have $b_r \geq \frac{t}{2}$ (this
can be verified by substituting the definition of $b_r$).
Therefore, $\|P_{S_{x,1}}\ket{\psi''_{all}}\| \geq 
\frac{\|\psi''_{all}\|^2}{K}$.

The balanced case is based on similar ideas but has some minor
changes in the expressions that appear in Lemma \ref{lem:project3-short}
(with the main change being $K\geq 20 N$ replaced by $K\geq 30Nd$,
which results in a bound of $c=\Omega(\frac{1}{Nd})$ instead of 
$c=\Omega(\frac{1}{N})$).
We discuss that in appendix \ref{app:recursion-full}.

{\bf Acknowledgments.}
I thank Richard Cleve and John Watrous for the discussion that lead
me to discovering the main idea of the paper.

\newpage
\begin{appendix}
\noindent
{\huge \bf Appendix}

\section{Lemmas on the structure of minimal certificates}
\label{app:structure}

\proof [of Lemma \ref{lem:structure}]
If $w$ belongs to $C$, then $C$ contains an extended minimal 
certificate for $T_w=0$ which, by the argument before the 
statement of lemma \ref{lem:structure} in section \ref{sec:structure}
(with $w$ instead of $v$) must contain an extended minimal
certificate for one of $T_{y_1}=0$ and $T_{y_2}=0$ and 
an extended minimal certificate for one of $T_{y_3}=0$ and 
$T_{y_4}=0$. 
From the construction of extended minimal certificates,
a certificate for $T_w=0$ contains $y_i$ if and only if
it contains a certificate for $T_{y_i}=0$.
\qed

\proof [of Lemma \ref{lem:0eigen}]
Since $H_v$ has non-zero entries only in the places corresponding to the edges of 
$T$, the amplitude of each basis state $\ket{u}$ in $H_v\ket{\psi_0}$ 
is just the sum of amplitudes of the neighbors of $u$ in $\ket{\psi_0}$,
multiplied by the appropriate factors. 
We need to show that, for every $u$, this sum is 0.

For vertices $u$ in the subtree $T_v$, if $u$ is at an even depth $2l$, then
its neighbors are at an odd depth ($2l-1$ or $2l+1$) and their 
amplitudes are 0. If $u$ is at an odd depth $2l+1$, let $p$ be the parent of
$u$ and let $y_1, y_2$ be the two children of $u$. 
If $p$ does not belong to $C$, than none of $p$'s descendants belongs
to $C$ as well, including $y_1$ and $y_2$. Then, all the neighbors
of $u$ have amplitude 0 in $\ket{\psi_0}$. 
If $p\in C$, then by Lemma \ref{lem:structure}, exactly one
of $y_1$ and $y_2$ belongs to $C$. Assume that $y_1\in C$.
Let $\alpha$ be the amplitude of $p$. By equation (\ref{eq:pc}),
the amplitude of $y_1$ is $-\frac{\sqrt[4]{m_p}}{\sqrt[4]{4m_{y_1}}}\alpha$.
The amplitude of $u$ in $H_v\ket{\psi}$ is
\[ H_{up} \alpha - H_{uy_1} \frac{\sqrt[4]{m_p}}{\sqrt[4]{4m_{y_1}}}\alpha =
\sqrt[4]{\frac{m_p}{2m_u}} \alpha - \sqrt[4]{\frac{2 m_{y_1}}{m_u}} 
\frac{\sqrt[4]{m_p}}{\sqrt[4]{4m_{y_1}}}\alpha = 0.\]
\qed

\proof [of Lemma \ref{lem:norm}]

Part (a). By induction on the depth of $T_v$. If $C_v$ is a leaf,
then $m_v=1$ and $\|\psi_{C_v}\|\leq 2\sqrt{1}-1$.
For the inductive case, decompose $C_v=\{v\}\cup C_{y_1}\cup C_{y_2}$.
By eq. (\ref{eq:pc}) and the inductive assumption, we have
\[ \|\psi_{C_v}\|^2 \leq  1 + \frac{\sqrt{m_v}}{2\sqrt{m_{y_1}}} 
(2\sqrt{m_{y_1}}-1) + \frac{\sqrt{m_v}}{2\sqrt{m_{y_2}}} 
(2\sqrt{m_{y_2}}-1) \]
\[ \leq 2\sqrt{m_v} - \sqrt{m_v} \left( \frac{1}{2\sqrt{m_{y_1}}} +
\frac{1}{2\sqrt{m_{y_2}}} \right) + 1 = 2\sqrt{m_v}-1 ,\]
with the first inequality following from the inductive assumption,
the second inequality following by rearranging terms and the
third following from $m_{y_1}=m_{y_2}=\frac{m_v}{2}$
(since the tree is balanced).

Part (b). By induction on the depth of $T_v$. If $C_v$ is a leaf,
then $m_v=d_v=1$ and $\|\psi_{C_v}\|=1\leq 2\sqrt{1}$.

For the inductive case, decompose $C_v=\{v\}\cup C_{y_1}\cup C_{y_2}$.
By eq. (\ref{eq:pc}) and the inductive assumption, we have
\[ \|\psi_{C_v}\|^2 = 1 + \frac{\sqrt{m_v}}{2\sqrt{m_{y_1}}} 
\|\psi_{C_{y_1}}\|^2 + \frac{\sqrt{m_v}}{2\sqrt{m_{y_2}}} 
\|\psi_{C_{y_2}}\|^2 \]
\[ \leq  
1+ \frac{\sqrt{m_v}}{2\sqrt{m_{y_1}}} 2 \sqrt{m_{y_1} d_{y_1}} +
\frac{\sqrt{m_v}}{2\sqrt{m_{y_2}}} 2 \sqrt{m_{y_2} d_{y_2}} = 
1+ \sqrt{m_v d_{y_1}} + \sqrt{m_v d_{y_2}} \]
\[
\leq 1 + 2\sqrt{m_v (d_v-1)}
 \leq \frac{m_v}{d_v} + 2 \sqrt{m_v (d_v -1)} \leq 2 \sqrt{m_v d_v} ,\]
with the first inequality following from the inductive assumption,
the second inequality following from $d_{y_1}\leq d_v-1$, $d_{y_2}\leq d_v-1$,
the third inequality following from $d_v\leq m_v$ (the depth of a tree is
always at most the number of leaves) and the fourth inequality following
from $\sqrt{A}-\sqrt{A-1}=\frac{1}{\sqrt{A}+\sqrt{A-1}}\geq \frac{1}{2\sqrt{A}}$.
\qed

\proof [of Lemma \ref{lem:fixed}]
To prove $U_1\ket{\psi_{C_1, C_2}}=\ket{\psi_{C_1, C_2}}$, 
we need to show $H\ket{\psi_{C_1, C_2}}=0$. 
Consider the amplitude of $\ket{u}$ in $H\ket{\psi_{C_1, C_2}}$.
If $u$ belongs to $T_{v_1}$ or $T_{v_2}$, 
its amplitude in $H\ket{\psi_{C_1, C_2}}$ 
are 0 by Lemma \ref{lem:0eigen}.
If $u=v$, its amplitude in $H\ket{\psi_{C_1, C_2}}$ is
\[ H_{u v_1} \sqrt[4]{m_{v_1}} - H_{u v_2} \sqrt[4]{m_{v_2}} =
\frac{\sqrt[4]{m_u}}{\sqrt[4]{2 m_{v_1}}} \sqrt[4]{m_{v_1}} - 
\frac{\sqrt[4]{m_u}}{\sqrt[4]{2 m_{v_2}}} \sqrt[4]{m_{v_2}} = 0.\]
If $u$ is outside $T_{v}$, then it has no neighbors in $T_v$, except
for possibly $v$ itself. That means that all of $u$'s neighbors
have amplitude 0 in $\ket{\psi_{C_1, C_2}}$ and 
the amplitude of $\ket{u}$ in $H\ket{\psi_{C_1, C_2}}$ is 0.

For $U_2$, we have $U_2\ket{\psi_{C_1, C_2}}=\ket{\psi_{C_1, C_2}}$ 
because the only leaves $v$ with $\ket{v}$ having a non-zero amplitude 
in $\ket{\psi_{C_1, C_2}}$ are those for which the 
corresponding variable $x_i$ belongs to $C_1$ or $C_2$ and all the variables $x_i$ in 
a certificate $C_j$ have $x_i=0$ which means that $U_2\ket{v}=\ket{v}$.
\qed

\proof [of Lemma \ref{lem:odd}]
We prove the lemma for the general case.

The proof is 
by induction. For the base case, let $u$ be a vertex of depth 1
(i.e. at a distance 1 from a leaf). 
Then, $u$ is connected to a leaf $w$. Since $w$ is a leaf, $u$ is the 
only neighbor of $w$. This means that the amplitude of $w$ in 
$H\ket{\psi}$ is equal to $H_{uw} \alpha_u$. Since $H\ket{\psi}=0$
and $H_{uw}\neq 0$, it must be the case that $\alpha_u=0$.

For the inductive case, assume that $\alpha_u=0$ for vertices $u$ of 
depth $2i+1$, for $i\in\{0, \ldots, l-1\}$. 
Let $u$ be a vertex at the depth $2l+1$. Let $c$ be one of
the two children of $u$. Let $v_1$ and $v_2$ be the children of $c$.
Then, the amplitude of $c$ in $H\ket{\psi}$ is equal to 
$H_{uc} \alpha_u+H_{cv_1}\alpha_{v_1}+H_{cv_2}\alpha_{v_2}$ and it must be equal to 0.
By the inductive assumption, $\alpha_{v_1}=\alpha_{v_2}=0$.
Therefore, $\alpha_u=0$.

The proof for the vertices in the tail is similar, starting with the 
vertex that is adjacent to the end of the tail and proceeding inductively 
towards the root. 
\qed

\section{Proof of Lemma \ref{lem:easy}}
\label{app:easy}

Similar statements have been proven before (e.g. \cite{Szegedy})
but none of them has the exact form that we need.
Therefore, we include the proof for completeness.

Let $\ket{\psi}$ be an eigenvector of $U_2 U_1$ with an 
eigenvalue $\lambda$.
We consider the following possibilities:
\begin{enumerate}
\item
$U_1\ket{\psi}=\ket{\psi}$. Then, $\ket{\psi}$ is an eigenvector
of $U_2U_1$ if and only if it is an eigenvector of $U_2$.
We cannot have $U_2\ket{\psi}=\ket{\psi}$ because,
by the conditions of the lemma, 
$\|P_{(S_2)^{\perp}}\ket{\psi}\| > 0$.
Since all eigenvalues of $U_2$ are $\pm 1$,
this means that $U_2\ket{\psi}=-\ket{\psi}$
and $U_2 U_1\ket{\psi}=-\ket{\psi}$.
\item
$U_1\ket{\psi}=-\ket{\psi}$. Again, $\ket{\psi}$ is an eigenvector
of $U_2U_1$ if and only if it is an eigenvector of $U_2$.
We cannot have $U_2\ket{\psi}=-\ket{\psi}$ because,
then $\ket{\psi}$ would belong to $S_1\cap S_2$ and
the lemma assumes that $S_1\cap S_2=\{\zv\}$.
Therefore, $U_2\ket{\psi}=\ket{\psi}$ and 
$U_2 U_1\ket{\psi}=-\ket{\psi}$.
\item
$U_1\ket{\psi}\neq \ket{\psi}$ and $U_1\ket{\psi}\neq -\ket{\psi}$.

Then, $\ket{\psi}$ is not an eigenvector of $U_1$.
This means that $\ket{\psi}$ and $\ket{\psi'}=U_1\ket{\psi}$
span a two dimensional subspace which we denote $\H_2$.
Since $U_1^2=I$, we also have $\ket{\psi}=U_1\ket{\psi'}$. 
This means that $U_1$ maps $\H_2$ to itself.
$U_2$ also maps $\H_2$ to itself, because it maps
$\ket{\psi'}$ to $U_2\ket{\psi'}=U_2U_1\ket{\psi'}=\lambda\ket{\psi}$
and, since $U_2^2=I$, this means that
$U_2\ket{\psi}=\lambda^{-1}\ket{\psi'}$.

Therefore, $U_2$ and $U_1$ both map $\H_2$ to itself.
Let $\ket{\psi_{i1}}$, $\ket{\psi_{i2}}$ be the eigenvectors of
$U_i$ in $\H_2$. One of $\ket{\psi_{11}}$, $\ket{\psi_{12}}$ must 
have an eigenvalue that is +1 and the other must have an eigenvalue
-1. (Otherwise, all of $\H_2$, including $\ket{\psi}$, 
would be eigenvectors of $U_1$ with the same eigenvalue and then 
we would have one of the first two cases.) Similarly, 
one of $\ket{\psi_{21}}$, $\ket{\psi_{22}}$ must 
have an eigenvalue +1 and the other must have an eigenvalue -1.

For simplicity, assume that $\ket{\psi_{11}}$ and $\ket{\psi_{21}}$
are the eigenvectors with eigenvalue 1.
Then, $U_2U_1$ is a composition of reflections w.r.t. 
$\ket{\psi_{11}}$ and $\ket{\psi_{21}}$.
By the analysis in \cite{Aha99}, the eigenvalues of $U_2U_1$ on $\H_2$
are $e^{\pm i\beta}$, where $\beta$ is the angle between 
$\ket{\psi_{11}}$ and $\ket{\psi_{21}}$.
We have 
\[ \|P_{(S_2)^{\perp}} \ket{\psi_{11}}\|^2= 
|\lbra \psi_{11} | \psi_{22} \rket|^2 = \sin^2 \beta .\]
By the conditions of the lemma, we have $\sin^2\beta\geq \epsilon$
which implies $\beta\in [\sqrt{\epsilon}, \frac{\pi}{2}]$.
\end{enumerate}

\section{Evaluating balanced trees: proof of Lemma \ref{lem:project3-short}}
\label{app:recursion-short}

Since
$a_v$ and $b_v$ only depend on $k$,
we will denote them $a_k$ and $b_k$. 
We first state some simple bounds on $a_k$ and $b_k$.

\begin{Claim}
\label{claim:ab}
\begin{enumerate}
\item[(a)]
$a_k\leq 1.1 \cdot 2^{k+1}$;
\item[(b)]
$b_k \geq 0.9 \cdot \frac{K}{2^k}$;
\item[(c)]
$a_k \leq 0.13 b_k$;
\end{enumerate}
\end{Claim}

\proof
The first two parts follow from 
$2\frac{2^{2k}}{K}\leq 2\frac{N}{20 N}=0.1$.
The third part follows by
\[ a_k \leq 1.1 \cdot 2^{k+1} \leq 0.11 \frac{K}{2^k} \leq 0.13 b_k ,\]
with the second inequality using $K\geq 20N\geq 20 \cdot 2^{2k}$
and the third inequality using part (b). 
\qed

The proof of Lemma \ref{lem:project3-short} is by an induction on $k$. 
The basis case is $k=0$. 
Then, the tree consists of the vertex $v$ only. The only possible
states $\ket{\psi}$ are multiples of $\ket{v}$. 
$v$ is also the only leaf, carrying a variable $x_1$ and $T_0(x_1)=x_1$.
If $x_1=0$, then $S_{x, 1}$ is empty, meaning that $L(\psi)=1$.
If $x_1=1$, then $S_{x, 1}$ consists of all multiples of $\ket{v}$,
meaning that $P_{S_{x, 1}} \ket{\psi}=\ket{\psi}$ and
$L(\psi)=-(K-1)$. In both cases, the lemma is true.

For the inductive case, let $z_1$ and $z_2$ be the children of $v$,
$y_1$ and $y_2$ be the children of $z_1$ and $y_3, y_4$ be the
children of $z_2$. 
We can decompose the state $\ket{\psi}$  as 
\[ \ket{\psi}=\alpha_v\ket{v}+ \ket{\psi_1} + \ket{\psi_2} \]
where $\ket{\psi_i} \in S_{z_i, 0}$.
We claim

\begin{Claim}
\label{claim:halfstep-short}
For every $i\in\{1, 2\}$, there exists $\ket{\psi'_i} \in S'_{z_i, 0}$
such that, for the state $\ket{\psi''_i}=\ket{\psi_i}+\ket{\psi'_i}$, we have
\begin{enumerate}
\item
If $T_{z_1}(x_1, \ldots, x_N)=0$, then 
\begin{equation}
\label{eq:half0-short}
 \| \psi''_i\| - \| P_{x, 1} \ket{\psi''_i} \| \leq	
\frac{-b_{k-1}}{2} |\alpha_v|^2.
\end{equation}
\item
If $T_{z_1}(x_1, \ldots, x_N)=1$, then 
\begin{equation}
\label{eq:half1-short}
 \| \psi''_i\| - \| P_{x, 1} \ket{\psi''_i} \| \leq	
\frac{a_{k}-1}{2} |\alpha_v|^2.
\end{equation}
\end{enumerate}
\end{Claim}

\proof
For typographical convenience, let $i=1$.
For the first part, $T_{z_1}=0$ if and only if $T_{y_1}=T_{y_2}=1$. 

By Lemma \ref{lem:odd}, the amplitude of $\ket{z_i}$ in $\ket{\psi_1}$
is 0. Therefore, we can decompose
\[ \ket{\psi_1}= \ket{\varphi_1}+\ket{\varphi_2}, \]
with $\ket{\varphi_i} \in S_{y_i, 0}$.
By the inductive assumption, there exist states
$\ket{\varphi'_1}, \ket{\varphi'_2}\in S'_{y_i, 0}$ such that,
for the states $\ket{\varphi''_i}=\ket{\varphi_i}+\ket{\varphi'_i}$, we have
\begin{equation}
\label{eq:ind1-short} 
\|\varphi''_i\|^2 - \| P_{S_{x, 1}} \ket{\varphi''_i}\|^2 
\leq -b_{k-1} |\alpha_{y_i}|^2 ,
\end{equation}
with $\alpha_{y_i}$ being the amplitude of $y_i$ in $\ket{\varphi_i}$.
We define $\ket{\psi'_1}=\ket{\varphi'_1}+\ket{\varphi'_2}$
and $\ket{\psi''_1}=\ket{\psi_1}+\ket{\psi'_1}$.
By summing equations (\ref{eq:ind1-short}) for $i=1$ and $i=2$, we get
\begin{equation}
\label{eq:case4-short} 
\|\psi''_1\|^2 - \| P_{S_{x, 1}} \ket{\psi''_1} \|^2 \leq 
-b_{k-1} \sum_{i=1}^2 |\alpha_{y_i}|^2 .
\end{equation}
Because of $\alpha_{y_1}+\alpha_{y_2}+\alpha_v=0$, we have
$|\alpha_{y_1}|^2+|\alpha_{y_2}|^2 \geq \frac{|\alpha_v|^2}{2}$.
Therefore, (\ref{eq:case4-short}) implies
\[ \|\psi''_1\|^2 - \| P_{S_{x, 1}} \ket{\psi''_1}\|^2 \leq 
-\frac{b_{k-1}}{2} |\alpha_{v}|^2 .\]

For the second part, we consider two cases:
\begin{enumerate}
\item
$y_1=y_2=0$.

Let $C$ be a minimal 0-certificate for $r=1$ and $\ket{\psi_{C}}$ 
be the state corresponding to this certificate.
Let $\ket{\psi_1}=\sum_u \alpha_u \ket{u}$. We define 
\[ \ket{\phi_1}=\frac{\alpha_{y_1}-\alpha_{y_2}}{2} \ket{\psi_{C}} .\]
Let $\ket{\psi_1}+\ket{\phi_1}=\sum_v\beta_v \ket{v}$.
Then, $\beta_{y_1}=\beta_{y_2}=\frac{\alpha_{y_1}+\alpha_{y_2}}{2}$.
Because of $\alpha_{y_1}+\alpha_{y_2}+\alpha_v=0$, we have
$\beta_{y_1}=\beta_{y_2}=-\frac{\alpha_v}{2}$.

We decompose
\[ \ket{\psi_1}+\ket{\phi_1} =\ket{\varphi_1}+\ket{\varphi_2} ,\]
with $\ket{\varphi_i}$ being a superposition over $T_{y_i}$.
By the inductive assumption, there exist states
$\ket{\varphi'_1}, \ket{\varphi'_2}\in S_{H, 0}$ such that,
for the states $\ket{\varphi''_i}=\ket{\varphi_i}+\ket{\varphi'_i}$, we have
\begin{equation}
\label{eq:ind2-short} 
\|\varphi''_i\|^2 - \| P_{S_{x, 1}} \ket{\varphi''_i}\|^2 
\leq a_{k-1} |\alpha_{y_i}|^2 .
\end{equation}
We define $\ket{\psi'_1}=\ket{\phi_1}+\ket{\varphi'_1}+\ket{\varphi'_2}$
and $\ket{\psi''_1}=\ket{\psi_1}+\ket{\psi'_1}$.
Summing up eq. (\ref{eq:ind2-short}) for $i=1, 2$ gives
\[
\|\psi''_1\|^2 - \| P_{S_{x, 1}} \ket{\psi''_1}\|^2 
\leq a_{k-1} (|\beta_{y_1}|^2+|\beta_{y_2}|^2) =
\frac{a_{k-1}}{2} |\alpha_{v}|^2 .
\]
The claim now follows from $a_{k-1}\leq a_k -1$ which is easy to prove.
\item
one of $y_1, y_2$ is 0 and the other is 1.

For typographical convenience, assume that $y_1=0$, $y_2=1$.
Once again, we decompose
\[ \ket{\psi_1}= \ket{\varphi_1}+\ket{\varphi_2}, \]
with $\ket{\varphi_i}$ being a superposition over $T_{y_i}$.
By the inductive assumption, there exist states
$\ket{\varphi'_1}, \ket{\varphi'_2}\in S_{H, 0}$ such that,
for the states $\ket{\varphi''_i}=\ket{\varphi_i}+\ket{\varphi'_i}$, we have
(\ref{eq:ind1-short}) for $i=1$ and (\ref{eq:ind2-short}) for $i=2$.
We define $\ket{\psi'_1}=\ket{\varphi'_1}+\ket{\varphi'_2}$,
$\ket{\psi''_1}=\ket{\psi_1}+\ket{\psi'_1}$ and
sum up the equations from the inductive assumption. This gives us
\begin{equation}
\label{eq:case3-short} 
 \|\psi''_1\|^2 - \| P_{S_{x, 1}} \ket{\psi''_1}\|^2 \leq 
 a_{k-1} |\alpha_{y_1}|^2 - b_{k-1} |\alpha_{y_2}|^2 .
\end{equation} 

Let $|\alpha_{y_2}|=\delta |\alpha_v|$. 
Then, because of $\alpha_{y_1}+\alpha_{y_2}+\alpha_v=0$, we have
$|\alpha_{y_1}|\leq (1+\delta) |\alpha_v|$.
We now upper-bound the expression
\[
a_{k-1} |\alpha_{y_1}|^2 - b_{k-1} |\alpha_{y_2}|^2 \leq 
a_{k-1} (1+\delta)^2 |\alpha_{v}|^2 - b_{k-1} \delta^2 |\alpha_{v}|^2 .
\]
Let $f(\delta)= a_{k-1} (1+\delta)^2 - b_{k-1} \delta^2$.
Then, $f'(\delta)= 2(1+\delta) a_{k-1}- 2\delta b_{k-1}$.
The maximum of $f(\delta)$ is achieved when $f'(\delta)=0$ which is 
equivalent to $\delta (a_{k-1} - b_{k-1})= - a_{k-1}$ and
$\delta = \frac{a_{k-1}}{b_{k-1}-a_{k-1}}$.
Then, 
\[ f(\delta) = a_{k-1} \left(\frac{b_{k-1}}{b_{k-1}-a_{k-1}}\right)^2 -
b_{k-1} \left(\frac{a_{k-1}}{b_{k-1}-a_{k-1}}\right)^2 =
 \frac{a_{k-1} b_{k-1}}{b_{k-1}-a_{k-1}} .\]
This means that 
\[ \|\psi''_1\|^2 - \| P_{S_{x, 1}} \ket{\psi''_1}\|^2  \leq 
\frac{a_{k-1} b_{k-1}}{b_{k-1}-a_{k-1}} |\alpha_v|^2 .\]
To complete the case, it suffices to show
\begin{equation}
\label{eq:we-need} 
2\frac{a_{k-1} b_{k-1}}{b_{k-1}-a_{k-1}}+1 \leq a_k .
\end{equation}
We have
\[ 2\frac{a_{k-1} b_{k-1}}{b_{k-1}-a_{k-1}} = 2 \left(1 + 
\frac{a_{k-1}}{b_{k-1}-a_{k-1}} \right) a_{k-1} \leq 
2 \left(1+\frac{1.1 \cdot 2^{k}}{0.87 b_k} \right) a_{k-1} \]
\[ \leq 2\left(1+\frac{1.1 \cdot 2^k}{0.87\cdot 0.9 \frac{K}{2^{k-1}}}\right)
a_{k-1} \leq 2 \left(1+1.41 \frac{2^{2k-1}}{K} \right) a_{k-1} \]
\begin{equation}
\label{eq:we-have} 
\leq 2 \left(1+2.82 \frac{2^{2k-2}}{K}\right) \left(
1+2\frac{2^{2k-2}}{K} \right) (2^{k}-1) \leq 
\left( 1+ 2\frac{2^{2k}}{K} \right)  (2^{k+1}-2) ,
\end{equation}
with the first inequality following from parts (a) and (c) of 
Claim \ref{claim:ab}, the second inequality following from part (b)
of Claim \ref{claim:ab}, 
the fourth inequality following by writing out $a_{k-1}$
and the last inequality following from 
$(1+2\delta)(1+2.82 \delta) \leq 1+8 \delta$ 
(where $\delta=\frac{2^{2k-2}}{K}$)
being true for sufficiently small $\delta$.
The equation (\ref{eq:we-need}) now follows by adding 1 to both sides 
of eq. (\ref{eq:we-have}).
\end{enumerate}
\qed

To deduce lemma \ref{lem:project3-short} from 
claim \ref{claim:halfstep-short}, we define
\[ \ket{\psi'}=\ket{\psi'_1}+\ket{\psi'_2} .\]
Let $\ket{\psi''}=\ket{\psi}+\ket{\psi'}$. 
Then, we also have
\[ \ket{\psi''}=\alpha_v \ket{v}+ \ket{\psi''_1}+\ket{\psi''_2} .\]

If $T_r=0$, then $T_{z_1}=T_{z_2}=1$. By summing up eq. (\ref{eq:half1-short}) 
for $i=1, 2$ and adding $|\alpha_v|^2$ to both sides, we get
\[ \| \psi''\| - \| P_{x, 1} \ket{\psi''} \| \leq a_{k} |\alpha_v|^2.\]

If $T_r=1$, then we again have two cases:
\begin{enumerate}
\item
$T_{z_1}=T_{z_2}=1$.

By summing up eq. (\ref{eq:half0-short}) for $i=1, 2$ and adding 
$|\alpha_v|^2$ to both sides, we get
\[ \| \psi''\| - \| P_{x, 1} \ket{\psi''} \| \leq - (b_{k-1}-1) |\alpha_v|^2.\]
The lemma follows from $b_{k-1}-1\geq b_k$ which is easy to prove.
\item
one of $z_1, z_2$ is 0 and the other is 1.

For typographical convenience, assume that $z_1=0$, $z_2=1$.
In this case, claim \ref{claim:halfstep-short} gives us 
\[ \| \psi''\| - \| P_{x, 1} \ket{\psi''} \| \leq - \frac{b_{k-1}-a_k-1}{2} 
|\alpha_v|^2.\]
To complete the proof, we need to show that $\frac{b_{k-1}-a_k-1}{2}\geq b_k$.
This follows by substituting the expressions for $a_{k}, b_{k-1}$ and $b_k$.
\end{enumerate}

\section{General case}
\label{app:recursion-full}

The counterpart of Lemma \ref{lem:project3-short} is

\begin{Lemma}
\label{lem:project3-full}
Let $K=30 N d$. 
For $v\in T$, define 
$\delta_v= \frac{5 m_v \sqrt{d_v}}{K} + \frac{d_v}{\sqrt{K}}$,
where $d_v$ is the depth of the subtree $T_v$. 
For any $v$ of even depth and any state $\ket{\psi}\in S_{v, 0}$, 
there exists a state $\ket{\psi'}\in S'_{v, 0}$ such that, 
for $\ket{\psi''}=\ket{\psi}+\ket{\psi'}$, we have 
\begin{enumerate}
\item
If $T_v(x_1, \ldots, x_N)=0$, then 
\[ L_v(\psi)\leq a_v, \mbox{~~} a_v=
2 ( 1+ \delta_v ) \sqrt{d_v m_v}  .\]
\item
If $T_k(x_1, \ldots, x_N)=1$, then 
\[ L_v(\psi)\leq  -b_v, \mbox{~~} b_v=
( 1- \delta_v) \frac{K}{\sqrt{m_v}}.\]
\end{enumerate}
\end{Lemma}

Observe that, because of $m_v \leq N$ and $d_v\leq d\leq N$, we always have
\[ \delta_v \leq \frac{5 N \sqrt{d}}{30 N d}+ \frac{d}{30 \sqrt{d N}} \leq
\frac{5}{30}+\frac{1}{30} = \frac{1}{5} .\]

\proof
By induction on the depth $l$ of the subtree $T_v$. The basis case is $l=0$. 
Then, the tree consists of $v$ only. The only possible
states $\ket{\psi}$ are multiples of $\ket{v}$. 
The root is also the only leaf, carrying a variable $x_i$ and $T_v(x_i)=x_i$.
If $x_i=0$, then $S_{x, 1}$ is empty, meaning that $L_v(\psi)=1$.
If $x_i=1$, then $S_{x, 1}$ consists of all multiples of $\ket{r}$,
meaning that $P_{S_{x, 1}} \ket{\psi}=\ket{\psi}$ and
$L_v(\psi)=-(K-1)$. In both cases, the lemma is true.

For the inductive case, let $z_1$ and $z_2$ be the children of $v$,
$y_1$ and $y_2$ be the children of $z_1$ and $y_3, y_4$ be the
children of $z_2$. 
We can decompose the state $\ket{\psi}$  as 
\begin{equation}
\label{eq:dec} 
\ket{\psi}=\alpha_v \ket{v}+ \ket{\psi_1} + \ket{\psi_2} 
\end{equation}
where $\ket{\psi_i}$ is a superposition over $\ket{u}, u\in T_{z_i}$.
We claim

\begin{Claim}
\label{claim:halfstep}
For every $i\in\{1, 2\}$, there exists $\ket{\psi'_i}\in S'_{z_i, 0}$, 
such that,
for the state $\ket{\psi''_i}=\ket{\psi_i}+\ket{\psi'_i}$, we have
\begin{enumerate}
\item
If $T_{z_1}(x_1, \ldots, x_N)=0$, then 
\begin{equation}
\label{eq:half0}
 \| \psi''_i\|^2 - \| P_{x, 1} \ket{\psi''_i} \|^2 \leq	
\frac{-b_{y_i}}{\sqrt{2}} |H_{v y_i}\alpha_v|^2.
\end{equation}
\item
If $T_{z_1}(x_1, \ldots, x_N)=1$, then 
\begin{equation}
\label{eq:half1}
 \| \psi''_i\|^2 - \| P_{x, 1} \ket{\psi''_i} \|^2 \leq 
\frac{a_{y_i}}{\sqrt{2}} 
 |H_{v y_i}\alpha_v|^2.
\end{equation}
\end{enumerate}
\end{Claim}

\proof
For typographical convenience, let $i=1$.
By Lemma \ref{lem:odd}, the amplitude of $z_1$ in $\ket{\psi_1}$ is 0.
Therefore, we can decompose
\[ \ket{\psi_1}= \ket{\varphi_1}+\ket{\varphi_2}, \]
with $\ket{\varphi_i}$ being a superposition over $T_{y_i}$.

For the first part, $T_{z_1}=0$ if and only if $T_{y_1}=T_{y_2}=1$. 
By the inductive assumption, there exist states
$\ket{\varphi'_1}\in S'_{y_1, 0}$, $\ket{\varphi'_2}\in \in S'_{y_2, 0}$ such that,
for the states $\ket{\varphi''_i}=\ket{\varphi_i}+\ket{\varphi'_i}$, we have
\begin{equation}
\label{eq:ind1} 
\|\varphi''_i\|^2 - \| P_{S_{x, 1}} \ket{\varphi''_i}\|^2 
\leq - b_{y_i} |\alpha_{y_i}|^2 ,
\end{equation}
with $\alpha_{y_i}$ being the amplitude of $y_i$ in $\ket{\varphi_i}$
(which is the same as its amplitude in $\ket{\psi_1}$).
We define $\ket{\psi'_1}=\ket{\varphi'_1}+\ket{\varphi'_2}$ and
$\ket{\psi''_1}=\ket{\psi_1}+\ket{\psi'_1}$.
By summing equations (\ref{eq:ind1}) for $i=1$ and $i=2$, we get
\begin{equation}
\label{eq:case4} 
\|\psi''_1\|^2 - \| P_{S_{x, 1}} \ket{\psi''_1} \|^2 \leq 
- \sum_{i=1}^2 b_{y_i} |\alpha_{y_i}|^2 .
\end{equation}
We would like to upperbound the right-hand side of this equation.
From $H\ket{\psi}=0$, we have
\begin{equation}
\label{eq:hreq} 
H_{y_1 z_1}\alpha_{y_1}+H_{y_2 z_1}\alpha_{y_2}+H_{v z_1}\alpha_v=0 .
\end{equation}
Define $x_i=-\frac{H_{y_i z_1}\alpha_{y_i}}{H_{v z_1}\alpha_v}$. 
Then, by dividing both sides of (\ref{eq:hreq}) by $-H_{v z_1}\alpha_v$, 
we have $x_1+x_2=1$.
By expressing $\alpha_{y_i}$ in terms of $x_i$, we get
\[ - \sum_{i=1}^2 b_{y_i} |\alpha_{y_i}|^2 =
- | H_{v z_1}\alpha_v |^2 \sum_{i=1}^2 \frac{b_{y_i}}{H^2_{y_i z_1}} 
|x_i|^2 \]
\[ \leq  - | H_{v z_1}\alpha_v |^2 ( 1- \delta_{z_1} ) 
\sum_{i=1}^2 \frac{K \sqrt{m_{z_1}}}{\sqrt{2} m_{y_i}} |x_i|^2 ,\]
where the last inequality follows by writing out $b_{y_i}$
and $H_{y_i z_1}$ and applying 
$\delta_{y_i}\leq \delta_{z_1}$
(which is true because both the size and the depth of $T_{y_i}$ are
less than the size and the depth of $T_{z_1}$).
To complete the proof, it suffices to show that
\begin{equation}
\label{eq:x1x2} 
\frac{|x_1|^2}{m_{y_1}}+\frac{|x_2|^2}{m_{y_2}} \geq \frac{1}{m_{z_1}}, 
\end{equation}
subject to the constraint $x_1+x_2=1$.
The left hand side of (\ref{eq:x1x2}) is minimized when $x_1$ and $x_2$ 
are both real. (Otherwise, one can replace $x_1$ and $x_2$ by
$x'_1=\frac{|x_1|}{|x_1|+|x_2|}$ and $x'_2=\frac{|x_2|}{|x_1|+|x_2|}$
and this does not increase the left hand side.)
Therefore, we can find the minimum of the left hand side of (\ref{eq:x1x2})
by substituting $x_2=1-x_1$ and taking the derivative
of the left hand side. That shows that the left hand side is minimized
by $x_1=\frac{m_{y_1}}{m_{y_1}+m_{y_2}}$, 
$x_2=\frac{m_{y_2}}{m_{y_1}+m_{y_2}}$.
Then, it is equal to $\frac{1}{m_{y_1}+m_{y_2}}=\frac{1}{m_{z_1}}$.

For the second part, we consider two cases:
\begin{enumerate}
\item
$y_1=y_2=0$.

Let $C_1, C_2$ be extended minimal certificates for $T_{y_1}=0$
and $T_{y_2}=0$, respectively and
$\gamma=\frac{\alpha_{y_2}-\alpha_{y_1}}{\sqrt[4]{m_{y_1}}+\sqrt[4]{m_{y_2}}}$.
We define
\[ \ket{\tilde{\varphi}_1} = \ket{\varphi_1} + \gamma \sqrt[4]{m_{y_1}} \ket{\psi_{C_1}} .\]
We define $\ket{\tilde{\varphi}_2}$ similarly, with - sign in the front of 
$\sqrt{m_{y_2}} \ket{\psi_{C_2}}$. We then
apply the inductive assumption to the states $\ket{\tilde{\varphi}_i}$,
obtaining $\ket{\varphi'_1}, \ket{\varphi'_2}$ such that
for $\ket{\varphi''_i}=\ket{\tilde{\varphi}_i}+\ket{\varphi'_i}$, we have
\begin{equation}
\label{eq:ind0} 
\|\varphi''_i\|^2 - \| P_{S_{x, 1}} \ket{\varphi''_i}\|^2 \leq a_{y_i} 
|\beta_{y_i}|^2 ,
\end{equation}
where $\beta_{y_i}$ is the amplitude of $y_i$ in $\ket{\tilde{\varphi}_i}$. 
We define 
\[ \ket{\psi'_1} = \ket{\varphi'_1}+\ket{\varphi'_2}+\gamma \ket{\psi_{C_1, C_2}} \]
and let $\ket{\psi''_1}=\ket{\psi_1}+\ket{\psi'_1}$.
Then, by summing equations (\ref{eq:ind0}), we get
\begin{equation}
\label{eq:case00}
 \|\psi''_1\|^2 - \| P_{S_{x, 1}} \ket{\psi''_1}\|^2 \leq 
 a_{y_1} |\beta_{y_1}|^2 + a_{y_2} |\beta_{y_2}|^2 .
\end{equation}
By our choice, we have $\beta_{y_1}=\beta_{y_2}$. 
Since $H\ket{\psi}=0$ and $H\ket{\psi'}=0$, we have $H\ket{\psi''}=0$.
By writing out the amplitude of $\ket{z_1}$ in $H\ket{\psi''}$, we get
\[ H_{y_1z_1} \beta_{y_1}+H_{y_2z_1} \beta_{y_2}+H_{z_1 v}\alpha_v=0 .\]
Therefore,
\[ \beta_{y_1}=\beta_{y_2} = - \frac{H_{z_1 v}\alpha_v}{H_{y_1z_1}+H_{y_2z_1}} .\]
By substituting that into (\ref{eq:case00}), we get
\[  \|\psi''_1\|^2 - \| P_{S_{x, 1}} \ket{\psi''_1}\|^2 \leq 
 \frac{a_{y_1}+a_{y_2}}{(H_{y_1z_1}+H_{y_2z_1})^2} |H_{z_1 v}\alpha_v|^2 .\]
We now expand the coefficient of $|H_{z_1 v}\alpha_v|^2$:
\[ \frac{a_{y_1}+a_{y_2}}{(H_{y_1z_1}+H_{y_2z_1})^2} \leq 
2 ( 1 + \delta_{z_1} )
\frac{\sqrt{d_{y_1} m_{y_1}}+\sqrt{d_{y_2} m_{y_2}}}{
\left(\frac{\sqrt[4]{2 m_{y_1}}}{\sqrt[4]{m_{z_1}}} + 
\frac{\sqrt[4]{2 m_{y_2}}}{\sqrt[4]{m_{z_1}}} \right)^2} \]
\[ \leq
\frac{1}{\sqrt{2}} 2 ( 1 + \delta_{z_1} ) \sqrt{d_{z_1} m_{z_1}}
\frac{\sqrt{m_{y_1}}+\sqrt{m_{y_2}}}{(\sqrt[4]{m_{y_1}}+
\sqrt[4]{m_{y_2}})^2} \]
\[ \leq \frac{1}{\sqrt{2}} 2 ( 1 + \delta_{z_1} ) \sqrt{d_{z_1} m_{z_1}} .\]
\item
one of $y_1, y_2$ is 0 and the other is 1.

For typographical convenience, assume that $y_1=0$, $y_2=1$.
By the inductive assumption, there exist states
$\ket{\varphi'_1}, \ket{\varphi'_2}\in S_{H, 0}$ such that,
for the states $\ket{\varphi''_i}=\ket{\varphi_i}+\ket{\varphi'_i}$, we have
\begin{equation}
\label{eq:ind-mix} 
\|\varphi''_i\|^2 - \| P_{S_{x, 1}} \ket{\varphi''_i}\|^2 \leq  c |\alpha_{y_i}|^2 ,
\end{equation}
with $c=a_{y_1}$ for $i=1$ and $c=-b_{y_2}$ for $i=2$.
We define $\ket{\psi'_1}=\ket{\varphi'_1}+\ket{\varphi'_2}$ and
$\ket{\psi''_1}=\ket{\psi_1}+\ket{\psi'_1}$.
By summing the equations (\ref{eq:ind-mix}), we get
\begin{equation}
\label{eq:case3} 
 \|\psi''_1\|^2 - \| P_{S_{x, 1}} \ket{\psi''_1}\|^2 \leq 
 a_{y_1} |\alpha_{y_1}|^2 - b_{y_2} |\alpha_{y_2}|^2 .
\end{equation} 
We switch to variables $x_i=-\frac{H_{y_i z_1}\alpha_{y_i}}{
H_{vz_1}\alpha_v}$, obtaining
\[  \|\psi''_1\|^2 - \| P_{S_{x, 1}} \ket{\psi''_1}\|^2 \leq 
|H_{v z_1}\alpha_{v}|^2  \left( 
\frac{a_{y_1}}{H^2_{y_1 z_1}} |x_1|^2 - 
\frac{b_{y_2}}{H^2_{y_2 z_1}} |x_2|^2 \right) .\]
Substituting the expressions for $a_{y_1}, b_{y_2}$,
$H_{y_1 z_1}$, $H_{y_2 z_1}$ gives
\[  \frac{a_{y_1}}{H^2_{y_1 z_1}} |x_1|^2 - 
\frac{b_{y_2}}{H^2_{y_2 z_1}} |x_2|^2 \]
\[ = \frac{1}{\sqrt{2}} \left(
(1+\delta_{y_1}) 2 \sqrt{d_{y_1} m_{z_1}} |x_1|^2 -
(1-\delta_{y_2}) \frac{K \sqrt{m_{z_1}}}{m_{y_2}} |x_2|^2 \right) \]
\begin{equation}
\label{eq:delta} 
\leq \frac{1}{\sqrt{2}} 2 (1+\delta_{y_1})
\sqrt{d_{y_1} m_{z_1}} \left( |x_1|^2 - \frac{K}{
3 \sqrt{d_{y_1}} m_{y_2}} |x_2|^2 \right) ,
\end{equation}
with the second inequality following by rearranging terms and applying 
$1-\delta_{y_2} \geq \frac{4}{5} \geq \frac{4}{6} (1+\delta_{y_1})$ which
is true because of $\delta_{y_1}\leq \frac{1}{5}$ 
and $\delta_{y_2}\leq\frac{1}{5}$.

Let $\delta=|x_2|$. Then $|x_1|\leq 1+\delta$ and the right hand side 
of (\ref{eq:delta}) is at most
\[ \frac{1}{\sqrt{2}} 2 (1+\delta_{y_1})
\sqrt{d_{y_1} m_{z_1}} \left( (1+\delta)^2 - 
\frac{K}{ 3 \sqrt{d_{y_1}} m_{y_2}} \delta^2 \right) .\]
Taking the derivative w.r.t. $\delta$ shows that
the expression in the brackets is maximized by 
\[ \delta=\frac{ 3 \sqrt{d_{y_1}} m_{y_2}}{K- 3 \sqrt{d_{y_1}} m_{y_2}},\]
in which case it is equal to $\frac{K}{K-3 \sqrt{d_{y_1}} m_{y_2}}$.
To complete the proof, we observe that
\[ (1+\delta_{y_1}) \frac{K}{K-3 \sqrt{d_{y_1}}m_{y_2}} =
1 + \delta_{y_1} + \frac{(1+\delta_{y_1}) 3\sqrt{d_{y_1}} m_{y_2}}
{K- 3\sqrt{d_{y_1}} m_{y_2}} \]
\[ \leq 1+\delta_{y_1} + 4 \frac{\sqrt{d_{y_1}} m_{y_2}}{K} 
< 1+ \delta_{y_1}+ 4\frac{\sqrt{d_{z_1}} m_{y_2}}{K} < 1+\delta_{z_1} ,\]
with the first inequality following from 
$\delta_{y_1}< \frac{1}{5}$ and $K-3\sqrt{d_{y_2}} m_{y_2}
\geq K - 3 N \sqrt{d} \leq \frac{27}{30}K$.
\end{enumerate}
\qed

To deduce lemma \ref{lem:project3-full} 
from claim \ref{claim:halfstep}, we define
\[ \ket{\psi'}=\ket{\psi'_1}+\ket{\psi'_2} .\]
Because of equation (\ref{eq:dec}), 
we can also express $\ket{\psi''}=\ket{\psi}+\ket{\psi'}$ as
\[ \ket{\psi''}=\alpha_v \ket{v}+ \ket{\psi''_1}+\ket{\psi''_2} .\]

If $T_r=0$, then $T_{z_1}=T_{z_2}=1$. By summing up eq. (\ref{eq:half1}) 
for $i=1, 2$ and adding $|\alpha_v|^2$ to both sides, we get
\begin{equation}
\label{eq:oldcase} 
\| \psi''\|^2 - \| P_{x, 1} \ket{\psi''} \|^2 \leq 
\left( \frac{a_{z_1}}{\sqrt{2}} H^2_{v z_1} + 
\frac{a_{z_2}}{\sqrt{2}}  H^2_{v z_2} + 1 \right) 
|\alpha_v|^2.
\end{equation}
By expanding $a_{z_i}$ and $H_{v z_i}$, we get
\[ \frac{a_{z_1}}{\sqrt{2}} H^2_{v z_1} + 
\frac{a_{z_2}}{\sqrt{2}}  H^2_{v z_2} + 1 \leq
( 1+ \delta_v ) 
\left( \frac{m_{z_1}\sqrt{d_{z_1}}}{\sqrt{m_{v}}} + 
\frac{m_{z_2}\sqrt{d_{z_2}}}{\sqrt{m_{v}}} 
+ 1 \right).\]
Since $m_v=m_{z_1}+m_{z_2}$, we have $\frac{m_{z_1}}{\sqrt{m_{v}}} + 
\frac{m_{z_2}}{\sqrt{m_{v}}}= \frac{m_v}{\sqrt{m_v}}=\sqrt{m_v}$. 
Together with $d_v=\max(d_{z_1}, d_{z_2})+1$, this implies
\[ \frac{m_{z_1}\sqrt{d_{z_1}}}{\sqrt{m_{v}}} + 
\frac{m_{z_2}\sqrt{d_{z_2}}}{\sqrt{m_{v}}} + 1 \leq
 \sqrt{m_v (d_v -1)}+1  \]
\[ \leq \sqrt{m_v (d_v -1)}+ \frac{\sqrt{m_v}}{\sqrt{d_v}} =
\sqrt{m_v} \frac{2\sqrt{d_v (d_v-1)}+1}{\sqrt{d_v}} \]
\[ \leq \sqrt{m_v} \frac{2 d_v}{\sqrt{d_v}} = 2\sqrt{m_v d_v} ,\]
with the second inequality following from $d_v \leq m_v$
(the depth of any tree is at most the number of leaves in it)
and the last inequality following from 
$\sqrt{d_v (d_v-1)}\leq d_v -\frac{1}{2}$.

If $T_r=1$, then we again have two cases:
\begin{enumerate}
\item
$T_{z_1}=T_{z_2}=1$.

By summing up eq. (\ref{eq:half0}) for $i=1, 2$ and adding 
$|\alpha_v|^2$ to both sides, we get
\[ \| \psi''\|^2 - \| P_{x, 1} \ket{\psi''} \|^2 \leq 
- \left( \frac{b_{z_1}}{\sqrt{2}} H^2_{v z_1} + 
\frac{b_{z_2}}{\sqrt{2}}  H^2_{v z_2} - 1 \right) 
 |\alpha_v|^2.\]
By expanding $b_{z_i}$ and $H_{v z_i}$ and simplifying, we get
\[ \frac{b_{z_1}}{\sqrt{2}} H^2_{v z_1} + 
\frac{b_{z_2}}{\sqrt{2}}  H^2_{v z_2} - 1 \geq
(1-\delta_{z_1}) \frac{K}{\sqrt{m_v}} + 
(1-\delta_{z_2}) \frac{K}{\sqrt{m_v}} - 1 \]
\[ \geq 2 ( 1 - \delta_v ) \frac{K}{\sqrt{m_v}} - 1 >
( 1 - \delta_v ) \frac{K}{\sqrt{m_v}} .\]

\item
one of $z_1, z_2$ is 0 and the other is 1.

For typographical convenience, assume that $z_1=0$, $z_2=1$.
In this case, claim \ref{claim:halfstep} gives us 
\[ \| \psi''\|^2 - \| P_{x, 1} \ket{\psi''} \|^2 \leq 
 \left( - \frac{b_{z_1}}{\sqrt{2}} H^2_{v z_1} + 
\frac{a_{z_2}}{\sqrt{2}}  H^2_{v z_2} + 1 \right) 
 |\alpha_r|^2.\]
We have
\[ - \frac{b_{z_1}}{\sqrt{2}} H^2_{v z_1} + 
\frac{a_{z_2}}{\sqrt{2}}  H^2_{v z_2} + 1 \leq 
- (1- \delta_{z_1} )  \frac{K}{\sqrt{m_v}} +
2 (1+ \delta_{z_2}) \frac{m_{z_2}\sqrt{d_{z_2}}}{\sqrt{m_v}} + 1 .\]
To prove that this is at most $-b_{v}$, we need to show that
\begin{equation}
\label{eq:case6} 
2 (1+\delta_{z_2} ) \frac{m_{z_2}\sqrt{d_{z_2}}}{\sqrt{m_v}} + 1  \leq
( \delta_v- \delta_{z_1} )  \frac{K}{\sqrt{m_v}} .
\end{equation}
This follows from
\[ ( \delta_v- \delta_{z_1} )  \frac{K}{\sqrt{m_v}} =
\left( \frac{5 m_{v}\sqrt{d_v}}{K} + \frac{d_v}{\sqrt{K}}
- \frac{5 m_{z_1}\sqrt{d_{z_1}}}{K} -\frac{d_{z_1}}{\sqrt{K}} \right)  
\frac{K}{\sqrt{m_v}} \]
\[
\geq
\left(\frac{5 m_{z_2}\sqrt{d_v}}{K} + \frac{1}{\sqrt{K}} \right)  
\frac{K}{\sqrt{m_v}} \geq  \frac{5 m_{z_2}\sqrt{d_{z_2}}}{\sqrt{m_v}} + 
\frac{\sqrt{K}}{\sqrt{m_v}} \]
\[ \geq 2 (1+\delta_{z_2} ) \frac{m_{z_2}\sqrt{d_{z_2}}}{\sqrt{m_v}} + 1
.\]
The first equality follows by writing out $\delta_v$ and $\delta_{z_1}$,
the next inequality follows from $m_v=m_{z_1}+m_{z_2}$ and $d_v>d_{z_1}$
and the last inequality follows from $\delta_{z_2}<\frac{1}{5}$
and $m_v\leq N \leq \frac{K}{30}$.
\end{enumerate} 
This completes the proof of Lemma \ref{lem:project3-full}.
\qed

The general case of Lemma \ref{lem:project2} now follows from
Lemma \ref{lem:project3-full} in the same way as the balanced case
of Lemma \ref{lem:project2} followed from Lemma \ref{lem:project3-short}. 

{\bf Distinction between balanced and general case.}
There are two reasons why our bound for the general case
has $O(\sqrt{Nd})$ instead of $O(\sqrt{N})$ for the balanced case:
\begin{enumerate}
\item
$z_1=0$, $z_2=1$ case of the proof of Lemma \ref{lem:project3-full}
which requires having $\sqrt{d_v m_v}$ instead of $\sqrt{m_v}$.
(The rest of the proof of Lemma \ref{lem:project3-full}
would still work with $\sqrt{m_v}$.)
\item
For the $T=0$ case, lemma \ref{lem:norm} has $O(\sqrt{m_v d_v})$ in
the general case and $O(\sqrt{m_v})$ in the balanced case.
\end{enumerate}
\end{appendix}
\end{document}